\documentclass[twoside]{article}
\usepackage{fleqn,espcrc2}

\usepackage{graphicx}

\newcommand{\AmS}{{\protect\the\textfont2
  A\kern-.1667em\lower.5ex\hbox{M}\kern-.125emS}}

\title{Background Studies for the Neutral Current Detector Array in the Sudbury Neutrino Observatory}

\author{K.M. Heeger, P.J. Doe, S.R. Elliott, R.G.H. Robertson, M.W.E Smith, T.D. Steiger, J.F. Wilkerson\address{Department of Physics, 
        University of Washington, \\ 
        Box 351560, Seattle, WA 98195, USA}
        \thanks{The authors would like to acknowledge the support of the Nuclear Physics Laboratory at the University of Washington.}}
       
\begin{document}

\begin{abstract}
An array of $^3$He-filled proportional counters will be used in the Sudbury Neutrino Observatory to measure the neutral-current interaction of neutrinos and deuterium. We describe the backgrounds to this detection method.

\vspace{1pc}
\end{abstract}

\maketitle

\section{Introduction}

Using heavy water as its target the Sudbury Neutrino Observatory (SNO) is the first solar neutrino detector capable of determining both the flux of electron neutrinos and the total flux of all active neutrinos from the Sun. Photomultiplier tubes are used to detect the \v{Cerenkov} light associated with charged-current (CC) neutrino interactions. The total flux of solar neutrinos is determined from the neutral-current (NC) dissociation of the deuteron and the subsequent detection of the free neutron. 
The SNO collaboration is pursuing two completely independent methods of measuring the NC signal. One involves the addition of a salt to absorb the free neutron on $^{35}$Cl. The second method is the deployment of an array of discrete $^3$He-filled proportional counters \cite{SNONIM,NCD} which detect the neutrons  from the NC interaction via the $^3$He(n,p)$^3$H reaction.  

Photodisintegration of the deuteron, and alphas from the decay chains of $^{238}$U and $^{232}$Th in the proportional counters are the most significant backgrounds to the $^3$He method. Radioassay techniques and underground studies provide first results on the backgrounds in the Neutral-Current Detectors (NCD). An {\em in-situ} background test experiment has been designed to measure the photodisintegration background from the NCD array prior to its installation in SNO.

\section{Neutral-Current Detection via Neutron Capture on $^3$He}

Construction of an array of $^3$He-filled proportional counters is nearing completion \cite{NCD}. With an estimated efficiency of 37\%, the expected number of neutron capture events on the NCD array is about 1700 events/year at the flux predicted by standard solar models \cite{SNONIM}.  
The use of two independent methods for the detection of the NC and CC signal allows their separation in real time and will help to determine signal and background events simultaneously. 

\subsection{Proportional Counter Signals}

Neutrons from the NC interaction of neutrinos with the D$_2$O thermalize and then capture via $^3$He(n,p)$^3$H in the proportional counters. They produce a 573 keV proton and a 191 keV triton that leave ionization tracks in the proportional counters. 

A number of different backgrounds contribute to the overall event rate. About 1000-10000 alpha particles per day are expected from the decay chains of $^{232}$Th and $^{238}$U. They enter the detection volume from the wall of the proportional counters and create a continuous background that underlies the $^3$He(n,p)$^3$H peak. Pulse shape analysis of the digitized proportional counter signals can be used to distinguish between neutron capture, alphas, and other ionization events. 
%

The $^3$He gas contains tritium that is of concern if the concentration is high enough to lead to random coincidences and to pileup of tritium decay pulses. $^3$H decays deposit an average of 6 keV/event in the proportional counters. Low temperature purification techniques are used to reduce $^3$H in the $^3$He-fill gas \cite{NCD}. 

The sensitivity of the $^3$He counters and their stringent background criteria make it essential to minimize all spurious pulses including signals induced by high voltage. The phenomenon of microscopic surface discharges and techniques for their reduction have been described elsewhere \cite{IEEE}. 

Gamma rays with an energy greater than 2.2 MeV contribute to the photodisintegration background.

\section{Photodisintegration Background}

Gamma rays above the deuteron binding energy can break the deuteron in the heavy water apart and produce neutrons that are indistinguishable from the NC signal. In the SNO detector the main sources of the photodisintegration background are $^{238}$U and $^{232}$Th in the water, gammas from photomultipliers, and ($\alpha$, p$\gamma$) and ($\alpha$, n$\gamma$) reactions on the PMT's and their support structure. $^{238}$U, $^{223}$Th, and cosmogenically activated $^{56}$Co in the nickel bodies of the Neutral-Current Detectors add to the photodisintegration background. 

\v{Cerenkov} light from associated gammas and radioassay techniques can be used to estimate these backgrounds. Results from radioassays indicate that the neutron background from the NCD array will contribute about 130 neutrons/year in the main detector volume, i.e. about 2.8\% of the expected neutrons from the NC interaction. 

\subsection{In-Situ Background Measurement}

The photodisintegration background from the Neutral Current Detector array can be assessed with the { \em Construction Hardware In-Situ Monitoring Experiment (CHIME)} prior to the array's installation in SNO. Observation of Cerenkov light from the CHIME background source will allow a quantitative estimate of the NCD originated photodisintegration background. The CHIME source consists of seven close-packed Neutral-Current Detectors. The construction materials and procedures are identical to those in the NCD array.  The CHIME is negative buoyant and will be deployed along the central vertical axis of the SNO detector using a specifically designed winding mechanism. The expected deployment date for the CHIME background source is summer 2000. 

\section{Neutron Background to the NC Signal}

The backgrounds to the NC signal rate of 4600 n/yr expected in standard solar models can be summarized in terms of the total number of neutrons produced in the detector. The photodisintegration background contributes an estimated 1600 n/yr to the internal D$_{2}$O backgrounds. The neutron background due to spallation reactions with cosmic rays muons (about 6000 n/yr) can be completely discriminated by the Cerenkov signal associated with each muon. Reactor and terrestrial $\bar{\nu_{e}}$ do not contribute more than 50 n/yr. The NCD array itself produces less than 230 n/yr in the D$_2$0. The total detector-internal background to the NC signal is estimated to be less than 1900 n/yr \cite{dunmore}. Determination of the photodisintegration and the additional external backgrounds is therefore essential for an accurate measurement of the NC interaction rate.  

\end{document}